\documentclass[twocolumn,showpacs,preprintnumbers,amsmath,amssymb]{revtex4}

\usepackage{graphicx}
\usepackage{dcolumn} 
\usepackage{bm}

\begin{document}

\title{Thermoelectric properties of the layered Pd oxide R$_2$PdO$_4$ 
(R = La, Nd, Sm and Gd)
}

\author{S. Shibasaki}
\email{g01k0374@suou.waseda.jp}
\affiliation{Department of Applied Physics, Waseda University, 
Tokyo 169-8555, Japan}

\author{I. Terasaki}
\affiliation{Department of Applied Physics, Waseda University, 
Tokyo 169-8555, Japan\\
also CREST, Japan Science and Technology Agency, Tokyo 108-0075, Japan}

\begin{abstract}

We prepared polycrystalline samples of R$_2$PdO$_4$ (R = La, Nd, Sm and Gd) 
using a NaCl-flux technique.
The measured resistivity is of the order of 10$^3$-10$^4$~$\Omega$cm 
at room temperature, which is two orders of magnitude smaller than the 
values reported so far.
We further studied the substitution effects of Ce for Nd in Nd$_{1.9}$Ce$_{0.1}$PdO$_4$,
where the substituted Ce decreases the resistivity 
and the magnitude of the thermopower.
The activation energy gap of 70-80~meV and the effective mass of 15 
evaluated from the measured data are suitable for thermoelectric materials,
but the mobility of 10$^{-6}$~cm$^2$/Vs is much lower than 
a typical value of 1-10~cm$^2$/Vs for other thermoelectric oxides.

\end{abstract}

\maketitle

\section{\label{sec:level1}Introduction}

The thermoelectric properties of transition-metal oxides have 
attracted increasing attention year by year, 
since the good thermoelectric properties of 
the layered Co oxide Na$_x$CoO$_2$ were 
discovered \cite{199701}.
Although the layered Co oxides are promising candidates 
for p-type thermoelectric materials, \cite{200001,200101} 
various n-type oxides, \textit{e.g.}, (Zn$_{1-x}$Al$_x$)O, 
(ZnO)$_k$In$_2$O$_3$,  Sr$_{0.9}$Dy$_{0.1}$TiO$_3$, 
and Sr$_{0.9}$Y$_{0.1}$TiO$_3$ 
are not yet better than them. \cite{199601,199801,200301,200401}
Thus a new n-type thermoelectric oxide is desired for all-oxide 
thermoelectric devices.

Recently, Ca$_{1-x}$Li$_x$Pd$_3$O$_4$ turned out to be a good p-type 
thermoelectric material \cite{200302}, 
and Ca$_{1-x}$Bi$_x$Pd$_3$O$_4$ was found to be an n-type material\cite{ICT2004}. 
Ca$_{1-x}$Bi$_x$Pd$_3$O$_4$ is the first example of the n-type Pd oxide, 
and is interesting in the sense that electrons are doped in the $d_{x^2-y^2}$ band 
separated from the $d_{3z^2-r^2}$ band by a small energy gap of 0-10~meV.
One serious drawback against good thermoelectrics
was a small effective mass with a low mobility. \cite{mahan}
In addition, substitution of Bi for Ca cannot change 
the doping level so widely, because the solubility limit of Bi is at most 15~\%
that corresponds to only 5~\% of electrons per Pd.
Thus, we had to find another Pd oxide for n-type thermoelectric materials 
in which carrier concentration can be changed more widely.

Among various Pd oxides, we focused on R$_2$PdO$_4$, 
(R = La, Nd, Sm, and Gd)
which is isomorphic to Nd$_{2-x}$Ce$_x$CuO$_4$, 
a famous n-type high-temperature superconductor \cite{198901}.
The Pd$^{2+}$ ion of $4d^8$ is surrounded with four planar oxygen atoms, 
in which the $d_{x^2-y^2}$ orbital is expected to be
the lowest unoccupied state.
Through substitution of Ce for R, 
we expect that electrons can be doped in the $d_{x^2-y^2}$ band,
and let the sample be an n-type conductor.
As is similar to the CuO$_2$ plane in Nd$_{2-x}$Ce$_x$CuO$_4$, 
we expect that the PdO$_2$ plane would be metallic with a wide $pd\sigma^*$ band.

\section{Experiment}

Polycrystalline samples of R$_2$PdO$_4$ (R = La, Nd, Sm and Gd) 
and Nd$_{1.9}$Ce$_{0.1}$PdO$_4$ were prepared by a solid-state reaction 
using a NaCl-flux technique \cite{200302}.
Stoichiometric amounts of La$_2$O$_3$, Nd$_2$O$_3$, Sm$_2$O$_3$, 
Gd$_2$O$_3$, CeO$_2$ and PdO of 99.9~\% purity were thoroughly mixed.
NaCl was then added in a mass ratio of 2:1, 
thoroughly mixed again, and were fired at 1073~K for 24~h in air.
The product was finely ground, and NaCl was rinsed out in hot distilled water.
The powder was dried, palletized and sintered at 1073~K for 12~h in air.
The low sintering temperature of 1073~K prevented cations from 
evaporating from the pellets, which was verified through the composition 
analyses of Ca$_{1-x}$Li$_x$Pd$_3$O$_4$ previously. \cite{200302}
We did not check the oxygen content, but as shown below, 
the magnitude of the thermopower of R$_2$PdO$_4$ 
is 200-600~$\mu$V/K at 300~K, clearly indicating 
that they are almost undoped.
This implies that the formal valence of Pd is close to $2+$, 
and that the oxygen content is almost 4.
Thus we refer the samples by nominal compositions in the present paper.

The black-colored samples were characterized by powder x-ray diffraction 
(XRD) using Cu $K\alpha$ radiation from 2$\theta$ = 10 to 100~deg.
The resistivity was measured by a four-probe method, 
and the thermopower was measured by a steady method.
The magnetization measurement was done with an SQUID susceptometer 
(Quantum Design MPMS) from 6 to 300~K in a field $B_{\rm ex}$ of 0.1~T.

\section{Results and Discussion}

\begin{figure}[t]
  \begin{center}
   \includegraphics[width=8cm,clip]{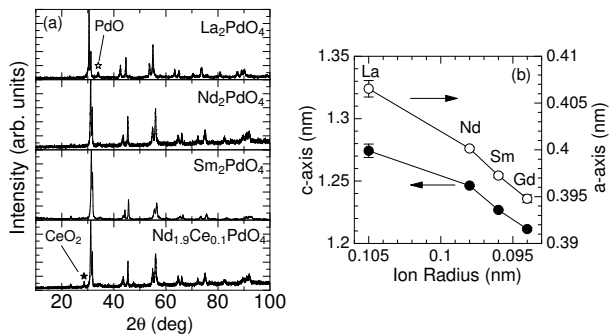}
   \caption{(a) X-ray diffraction patterns of
   R$_2$PdO$_4$ (R = La, Nd and Sm)
   and Nd$_{1.9}$Ce$_{0.1}$PdO$_4$,
   and 
   (b) lattice constants of R$_2$PdO$_4$ 
   (R = La, Nd, Sm and Gd). }
   \label{figxr}
  \end{center}
 \end{figure}

Figure~\ref{figxr}(a) shows typical XRD patterns of the prepared samples, 
which are indexed as R$_2$PdO$_4$ with a small amount 
of unreacted PdO and CeO$_2$.
In previous works \cite{198201,198902}, polycrystalline samples of 
La$_2$PdO$_4$ included impurity phases like La$_2$Pd$_2$O$_5$ or La$_4$PdO$_7$.
Figure~\ref{figxr}(a) shows no trace of such phases, which indicates 
that the NaCl-flux technique is very effective to stabilize La$_2$PdO$_4$.
Lattice constants evaluated from Fig.~\ref{figxr}(a) 
are plotted as a function of ionic radius of R in Fig.~\ref{figxr}(b), 
indicating that the $a$- and $c$-axis lengths systematically decrease with 
decreasing ion radius of R.

\begin{figure}[t]
  \begin{center}
   \includegraphics[width=6cm,clip]{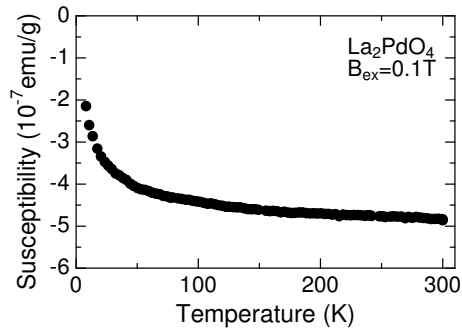}
   \caption{The susceptibility of La$_2$PdO$_4$ in an external field 
   $B_{\rm ex}$ of 0.1~T.}
   \label{figmag}
  \end{center}
 \end{figure}

In order to check the magnetic state of the Pd ion, we measured 
the susceptibility of La$_2$PdO$_4$. 
Note that Nd, Sm and Gd are magnetic ions with localized $f$ electrons, 
which will obscure the susceptibility of the Pd ions.
As shown in Fig.~\ref{figmag},
the susceptibility of La$_2$PdO$_4$ is as small as 
that of a diamagnetic substance like water.
This means that the susceptibility is determined 
by the diamagnetism of core electrons, 
and that Pd$^{2+}$ is nonmagnetic.
Thus we can safely assume that the Pd ions are
in the low-spin state with the configuration 
of $(t_{2g})^6(d_{3z^2-r^2})^2(d_{x^2-y^2})^0$,
and that the $d_{x^2-y^2}$ band is unoccupied, 
as expected in the introduction.
Toward 0~K, the susceptibility slightly increases, 
possibly caused by a small amount of impurity phases.
Fitting teh data with the Curie-Weiss law, we evaluated 
a maximum volume fraction of the magnetic impurity to be 0.1~\%.

\begin{figure}[t]
  \begin{center}
   \includegraphics[width=6cm,clip]{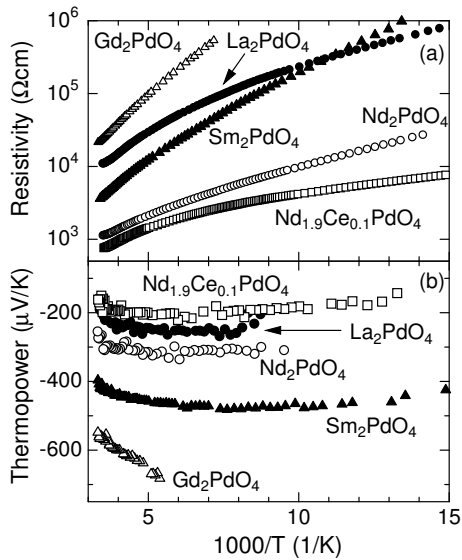}
   \caption{
   (a) Resistivity and (b) thermopower of R$_2$PdO$_4$ 
   (R = La, Nd, Sm and Gd) 
   and Nd$_{1.9}$Ce$_{0.1}$PdO$_4$.
   }
   \label{figrrs}
  \end{center}
 \end{figure}

Figure~\ref{figrrs}(a) shows the resistivity of the prepared samples.
The magnitudes are 10$^3$-10$^4$~$\Omega$cm at 300~K, 
and rapidly increases with decreasing temperature.
These results indicate that R$_2$PdO$_4$ is an undoped insulator,
which is consistent with the small diamagnetic susceptibility of La$_2$PdO$_4$.
We note that the resistivity  of our samples at room temperature 
is two orders of magnitude smaller than the previously reported 
values \cite{198202}.
This is consistent with the fact that our samples
did not include impurity phases like R$_2$Pd$_2$O$_5$ and R$_4$PdO$_7$.

Figure~\ref{figrrs}(b) shows that the thermopower of the prepared samples.
All the samples show a negative sign,
indicating that R$_2$PdO$_4$ is an n-type conductor.
A similar negative thermopower is seen in the n-type copper oxide
Nd$_2$CuO$_4$ that is isomorphic to the title compound. \cite{198901}
The magnitude is as large as 200-600~$\mu$V/K at 300~K, 
which indicates that the carriers are almost undoped.
Considering that conventional thermoelectric materials show a thermopower
of 200~$\mu$V/K at 300~K, we conclude that the carrier density is 
somewhat lower than that of them. 
Comparing with a typical carrier density 
(10$^{19}$-10$^{20}$~cm$^3$) of
conventional thermoelectric materials,
we estimate the carrier density of R$_2$PdO$_4$ 
to be of the order of 0.01 electrons per Pd ion.

One may notice that the magnitude of the thermopower 
systematically increases in going from R = La to Gd, which suggests that 
the carrier density systematically decreases from La to Gd.
Similarly, the resistivity is highest for Gd$_2$PdO$_4$, 
and is lowest for Nd$_2$PdO$_4$.
(La$_2$PdO$_4$ shows higher resistivity than Sm$_2$PdO$_4$.
This does not follow the trend, 
possibly owing to poorer sintering.
In fact, a larger amount of PdO is detected 
in La$_2$PdO$_4$ in Fig.~\ref{figxr}.)
Since the estimated carrier density is 
of the order of 1~\% per Pd, 
a small cation deficiency of less than 1~\% will 
significantly affect the transport parameters.
The ionic radius of R systematically decreases with increasing atomic number,
and the deficiency in the R site will possibly vary with the ionic radius.
A similar example is seen in the parent insulator of high-temperature 
superconductor Bi$_2$Sr$_2$Er$_{1-x}$Dy$_x$Cu$_2$O$_{8+\delta}$, in which 
the carrier density systematically decreases with increasing $x$. \cite{yanagi}

Let us analyze the resistivity and thermopower more quantitatively.
As mentioned above, R$_2$PdO$_4$ is a nearly undoped semiconductor 
with an energy gap between the $d_{x^2-y^2}$ and $d_{3z^2-r^2}$ bands.
Thus we expect a conventional semiconductor physics that 
the transport properties at high temperatures 
are of activation type characterized 
by the band gap, and those at low temperatures are 
of variable-range-hopping (VRH) type.
The crossover will occur
when the thermally activated carriers are comparable 
to the doped carriers in number.
That is, the more carriers are doped, 
the higher the crossover temperature is.
Accordingly, we expect that the activation-type conduction is 
realized below 300~K for the least doped sample Gd$_2$PdO$_4$.

The activation-type resistivity and thermopower are
written by
 \begin{eqnarray}
  \rho &\propto& \exp\left[\frac{E_g}{k_BT}\right],\\
  S &\propto& \left[\frac{E_s}{eT}+\mbox{const.} \right].
   \label{eqSact}
 \end{eqnarray}
Then the resistivity plotted in a log scale 
and the thermopower plotted in a linear scale are inversely proportional to temperature.
As shown in Fig.~\ref{figrrs}, the data for Gd$_2$PdO$_4$ roughly obey these relations.
The activation energies are evaluated to be $E_g$=80.7~meV from the resistivity
and $E_s$=73.6~meV from the thermopower.
A relation of $E_g\sim E_s$ strongly suggests that the activation 
process in Gd$_2$PdO$_4$ is determined by carrier density 
thermally excited across the band gap.
In other words, $E_g\sim E_s$ is the energy gap between 
the $d_{x^2-y^2}$ and $d_{3z^2-r^2}$ bands.
Since these two bands would be degenerate in the cubic symmetry, 
the small band gap less than 100~meV 
evaluated here is reasonable.

We should note here that the relation of $E_g\gg E_s$ 
is often seen in 3$d$ transition-metal oxides, where 
the carriers distort the lattice nearby through
the strong electron-phonon interaction to be stabilized 
in the potential well of the distorted lattice.\cite{kobayashi}
This ``self-trapped'' state is called a (small) polaron,
and such a material is called a polaronic conductor.
We expect an activation-type conduction in polaronic conductors, 
where the activation energy comes from the potential well
(the polaron binding energy), not from the band gap.
This means that the mobility or the scattering time obeys 
the activation law.
Since the thermopower is independent of the scattering time
in the lowest order approximation, \cite{197501}
$E_g \gg E_s$ is expected in a polaron semiconductor,
where $E_s$ is the band gap and $E_g-E_s$ is 
the polaron binding energy.\cite{palstra}
In this context, the relation $E_g\sim E_s$ in Gd$_2$PdO$_4$ is rather unusual 
in the transition-metal oxides, which implies that 
a band picture works well to some extent
in the 4$d$ transition-metal oxides, possibly owing to the broader $4d$
bands than the $3d$ bands.\cite{terra}

\begin{figure}[t]
  \begin{center}
   \includegraphics[width=6cm,clip]{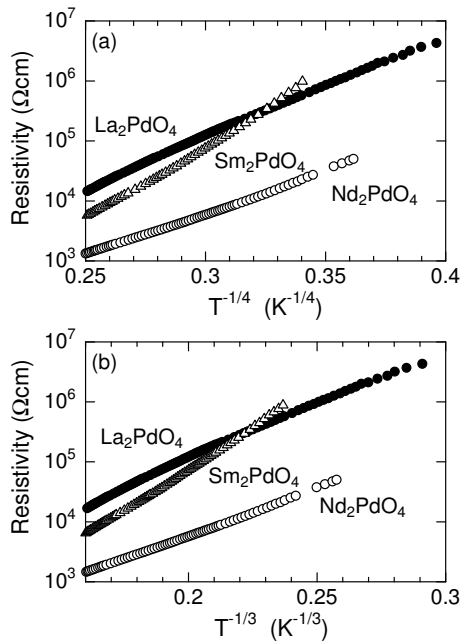}
   \caption{
   Resistivity for R$_2$PdO$_4$ (R = La, Nd and Sm)
   plotted as a function of (a) $T^{-1/4}$ 
   and (b) $T^{-1/3}$.
   }
   \label{figvrh}
  \end{center}
 \end{figure}

The other three samples (La$_2$PdO$_4$, Nd$_2$PdO$_4$ and Sm$_2$PdO$_4$)
are more doped than Gd$_2$PdO$_4$, and show the VRH-type transport
expressed by
 \begin{equation}
  \rho \propto \exp\left[ \left(\frac{T_0}{T}\right)^{\frac{1}{d+1}}\right],
 \end{equation}
where $d$ is the dimension of the system \cite{197901}.
The data are fitted well with $d=2$ or $d=3$, as shown in Fig.~\ref{figvrh}
The characteristic temperature $T_0$ is a measure of the localization 
length $\xi$ through $T_0 \propto 1/k_BD\xi^d$,
where $D$ is the density of states at the Fermi energy.
Figure~\ref{figvrh} shows that $T_0$ 
is roughly the same among R = La, Nd and Sm.
Since $D$ does not change very much with R, the localization length $\xi$ 
is roughly the same in R$_2$PdO$_4$.

The thermopower of the three samples 
shows a broad maximum around 100-200~K in Fig.~\ref{figrrs}(b), 
which is consistent with the VRH picture of a slightly doped semiconductor.
The VRH-type thermopower is given by \cite{198301}
 \begin{equation}
  S \propto T^{\frac{d-1}{d+1}}.
 \end{equation}
Accordingly, the sublinear thermopower 
($S\propto T^{1/3}$ for $d=2$ or $S\propto T^{1/2}$ for $d=3$)
is expected at low temperatures, and the activation-type 
thermopower of Eq.~(\ref{eqSact}) is expected at high temperatures.
The broad maximum observed around 100-200~K is a sign for the crossover between the two regimes. 
The sublinear thermopower should have been observed below 100~K, but
the sample resistance was too high to measure reliable thermoelectric 
voltage below 100~K in our experimental setup.

The thermoelectric performance is characterized by the power factor ($S^2/\rho$),
which is a measure of the maximum electric power generated from
thermal energy.\cite{mahan} 
Figure~\ref{figrrs} shows that $S^2/\rho$ at 300~K takes the maximum for R = Nd.
Thus we partially substituted Ce for Nd in Nd$_2$PdO$_4$ to see the doping effects.
Although a tiny peak of CeO$_2$ is present in $x$=0.1 in Fig.~\ref{figxr}(a), 
the resistivity and thermopower 
decrease from those for Nd$_2$PdO$_4$, as shown in Fig.~\ref{figrrs}.
This suggests that a substantial portion of Ce is substituted for Nd
to supply holes, and further suggests that Nd$_{1.9}$Ce$_{0.1}$PdO$_4$ is a metal
(or a degenerate semiconductor) with a finite Fermi energy $E_F$ in the
low temperature limit ($T \to 0$).
Then the thermopower at low temperatures is given by \cite{197501}
 \begin{equation}
  S=-\frac{\pi^2}{2e}\frac{k_B^2T}{E_F}.
   \label{eqS}
 \end{equation}
On the other hand, in the high temperature limit of $E_F \ll k_BT$,
the thermopower becomes temperature-independent, 
and obeys the Heikes formula given by\cite{197601}
 \begin{equation}
  \label{eqheikes}
  S=-\frac{k_B}{e}\ln\left|\frac{2(1-N)}{N}\right|,
 \end{equation}
where $N$ is the number of carriers per Pd.

 \begin{figure}[t]
  \begin{center}
   \includegraphics[width=6cm,clip]{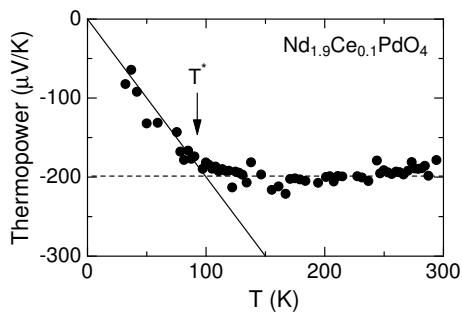}
   \caption{
   Thermopower of Nd$_{1.9}$Ce$_{0.1}$PdO$_4$.
   The solid and dotted lines represent Eqs. (\ref{eqS})
   and (\ref{eqheikes}), respectively.
   The crossover temperature $T^*$ is indicated by an arrow.
   }
   \label{fig5}
  \end{center}
 \end{figure}

As shown by the solid and dotted lines in Fig.~\ref{fig5}, 
the fitting with Eqs.~(\ref{eqS}) and (\ref{eqheikes}) are satisfactory.
$E_F$ is evaluated to be 16.5~meV from Eq.~(\ref{eqS}), 
corresponding to a thermal energy of 200~K that is close to the 
crossover temperature $T^* \sim$100~K.
Substituting the saturated value of $S=-$200~$\mu$V/K into 
Eq.~(\ref{eqheikes}), 
we get the carrier density $n$=1.6$\times10^{21}$~cm$^{-3}$, 
which is of the same order of the ideal value 
that each Ce ion supplies one electron. 
Combining with $n$ and $E_F$, we evaluate the effective mass $m^*$ 
to be 15 through the following equation
 \begin{equation}
  \label{eqef}
  E_F=\hbar^2\pi d_c\frac{n}{m^*},
 \end{equation}
where $d_c$ is the interlayer distance \cite{199001}.
The heavy $m^*$ and the small $E_g$ are favorable to good 
thermoelectrics \cite{mahan}, 
but the mobility $\mu$ evaluated from $\mu=1/ne\rho$ 
is of the order of 10$^{-6}$~cm$^2$/Vs at 300~K
which is much smaller than a typical mobility of thermoelectric oxides
(1-10~cm$^2$/Vs at 300~K).

\section{Summary}

We prepared polycrystalline samples of R$_2$PdO$_4$ 
(R = La, Nd, Sm and Gd) 
and Nd$_{1.9}$Ce$_{0.1}$PdO$_4$ 
using NaCl flux. 
They are n-type materials 
with the thermopower of $-$600$\sim$ $-$200~$\mu$V/K at room temperature.
The resistivity of R$_2$PdO$_4$ is semiconductive,
characterized by an activation energy of 70-80~meV.
By the Ce substitution, electrons are doped to reduce 
the resistivity and thermopower in magnitude.
From the thermopower, the carrier concentration and the effective mass 
are evaluated to be 1.6$\times 10^{21}$~cm$^{-3}$ and 15, respectively.
The mobility is poor, which might be improved through different synthetic routes.

\begin{acknowledgments}

We thank W. Kobayashi, S. Okada and S. Ishiwata for useful comments.

\end{acknowledgments}

\end{document}